# Customising Agent Based Analysis Towards Analysis of Disaster Management Knowledge

**Dedi Iskandar Inan**
Universitas Negeri Papua, Indonesia
School of Computing and Information Technology
University of Wollongong, Australia
dii740@uowmail.edu.au; d.inan@unipa.ac.id

**Ghassan Beydoun**
School of System, Management and Leadership
University of Technology Sydney, Australia
ghassan.beydoun@uts.edu.au

**Simon Opper**
Emergency Risk Management Branch
SES New South Wales, Australia
simon.opper@gmail.com

## Abstract

In developed countries such as Australia, for recurring disasters (e.g. floods), there are dedicated document repositories of Disaster Management Plans (DISPLANs), and supporting doctrine and processes that are used to prepare organisations and communities for disasters. They are maintained on an ongoing cyclical basis and form a key information source for community education, engagement and awareness programme in the preparation for and mitigation of disasters. DISPLANS, generally in semi-structured text document format, are then accessed and activated during the response and recovery to incidents to coordinate emergency service and community safety actions. However, accessing the appropriate plan and the specific knowledge within the text document from across its conceptual areas in a timely manner and sharing activities between stakeholders requires intimate domain knowledge of the plan contents and its development. This paper describes progress on an ongoing project with NSW State Emergency Service (NSW SES) to convert DISPLANs into a collection of knowledge units that can be stored in a unified repository with the goal to form the basis of a future knowledge sharing capability. All Australian emergency services covering a wide range of hazards develop DISPLANs of various structure and intent, in general the plans are created as instances of a template, for example those which are developed centrally by the NSW and Victorian SES's State planning policies. In this paper, we illustrate how by using selected templates as part of an elaborate agent-based process, we can apply agent-oriented analysis more efficiently to convert extant DISPLANs into a centralised repository. The repository is structured as a layered abstraction according to Meta Object Facility (MOF). The work is illustrated using DISPLANs along the flood-prone Murrumbidgee River in central NSW.

# 1   Introduction

Effective Disaster Management (DM) requires harnessing and sharing knowledge between stakeholders in a timely and user specific manner for effective preparation and reduction of the impact of a disaster. DM incorporates many elements within a 'Planning Process', a process which makes all parties involved in DM aware of the risks and their actions prior to an incident occurring. The planning process includes but is not limited to policy, doctrine and process adopted by organisations, which in Australia mainly comprises: 1. Emergency Risk Management (Australian Emergency Management Manuals Series (AEM Series), Manual 5, – identification and treatment of risk 2. Emergency Planning (AEM Series, Manual 43) – undertaking planning and documenting the arrangements to manage residual risk 3. Hazard mitigation Programs (AEM Series, Manuals 5, 6 and 19) – physical structures and modification of built and natural environments and resilience building measures to reduce hazard consequence. 4. Community education, awareness and engagement programs (AEM Series, Manual 45) – community resilience building. 5. Total Warning System (AEM Series, Manual 21). 6. Incident Management Frameworks – organisational principles, team structures and incident tasks and objective management (AFAC, Australasian Interagency Incident Management System, AIIMS, 2013), (AEM Series, Manuals 1, 17, 31), and 7. Recovery (AEM Series, Manual 29) – arrangements to return community functioning and provide welfare and support.  The knowledge breadth of these areas is immense in terms of societal facets, organisational effort and experience, and unsurprisingly the relevant flow of actions is often poorly understood by participants and especially the community.  The NSW SES prepares and maintains some 123 individual Local Flood Plans across NSW Local Government Areas, along with 4 overarching State level plans in support of the planning elements above. To effectively maintain this large number of plans a central template is used to structure the elements above which are then customised to local hazards, risks, strategies and community characteristics. Other flood emergency managers as well as other hazard managers, such as bushfire agencies, maintain similarly large numbers of Local and Regional level disaster plans dealing with various combinations of the elements above.

The outcome of planning these elements include physical mitigation such as building standards and modifications or physical measures such as flood levees, community education, engagement and awareness programs, response and rescue planning, developing strategies for evacuation and warning, and finally for recovery and rebuilding. Throughout these processes, knowledge, risk data, key actions and a complex structure of roles and responsibilities is generated across a large range of domains within DM. A DISPLAN is the documented output of this process for emergency organisations and may link and interact to multiple external DISPLANs. In the current paradigm, commonly an agency DISPLAN or a template derived from it will be the same plan used by communities. However, the inherent complexity and the amount of activities required in DM may in itself make navigating this knowledge complex for external stakeholders, especially the community and individuals. To support the key goal of emergency managers to deliver community focused disaster management which fosters resilient communities who understand their risks, have the capacity to manage them (National Strategy for Disaster Resilience, 2011), more timely, tailored and locally specific information derived from DISPLANs and other doctrine and process are required.

This research extends joint effort with the NSW State Emergency Service and in the future other emergency services in investigating opportunities to improve the outcomes of the planning process. Of the elements involved in the DM planning process e.g. risk assessment, warning or rescue, this work is focused on the overarching roles and responsibilities assigned within a DISPLAN which spans across all of the planning elements, rather than any specific domain areas. In this work, we adopt a DM unified metamodel, DMM, as the underlying representation that was developed in (Othman et al., 2014). The DMM enables layering of abstractions of various knowledge constructs giving free flow access to any point in DISPLANs. The challenge addressed in this paper is how to convert DISPLANs to concepts and notation from DMM. Specifically, the paper shows how existing DISPLAN templates can be harnessed in creating specialised agent-based models that can be more effective in the analysis of the instances of the DISPLANs.

Prior use of DISPLAN templates contributes to a better DISPLAN representation that can be effectively and efficiently understood. DISPLAN template for NSW SES for example have been continually evolving and revised by subject matter experts since the early 1990's. Thus the template serves as a robust classifier of knowledge representation that will be used to instantiate other DISPLAN based on the significant subject area expertise. Although not yet tested, it may be possible that the detailed structure of long established plans for one hazard, such as flood, may have a core set of domain structures that can be effectively applied to other hazards such as bushfire, and vice versa. In fact, the merging of multiple hazard plans may allow the derivation of a highly developed and more complete set of domain knowledge for reuse across all hazards. While hazard and agency and jurisdictional specific elements will exist, this hypothesis is supported by the approach taken by emergency services of implementing an all-hazard approach in their doctrine, or in other words interoperable arrangements, data dictionaries and organisational structures between the various organisations, facilitating coordination. The all- hazards paradigm also extends to community, who in preparing for and reducing the impact of one type of disaster are effectively becoming more resilient to a range of other hazards. Resilient communities are the goal of global disaster risk management policy and practice (UNISDR Resilient Cities Project 2012, Sendai Framework, 2015, Australian Commonwealth Government, National Strategy for Disaster Resilience, 2011) and in this context a goal of this work is to make information contained in the various repositories of hazard DISPLANs and community safety information along with public warnings, more locally specific and tailored by drawing these from the metamodel. By merging domain and metamodel data with temporal, spatial and social media data, more targeted information can be delivered to a much wider set of specific audiences and channels such as discrete linguistic communities or known high hazard risk areas and vulnerable groups. The paper deploys Agent-Based Models (ABMs) towards this end. ABMs lend themselves to representing organisational know-how and DM processes. They emphasise the constructs of roles, agents and organisations to represent systems behaviours. Much know-how and processes in DISPLANs across Australia are actually expressed in such terms. With appropriate supporting tools, this knowledge can be deposited and shared using a DMM-based system.

The rest of this paper is organised as follows: the next section reviews the background and related work. The third section presents our knowledge analysis framework to customize and generate a particular DISPLAN and subsequently convert it into the DMM constructs. The fourth section illustrates the approach using an actual SES NSW DISPLAN template. Finally, the paper concludes with a discussion of future work.

## 2　Related work

In Australia, the PPRR (Prevention, Preparedness, Response, Recovery) model is typically used to organise DM knowledge (Rogers, 2011). Various DM activities and knowledge units required throughout the DM processes are organised according to the sequence of the four phases: Prevention (P), Preparedness (P), Response (R) and Recovery (R). DISPLAN may be focused on any combination of these phases depending on the intent and audience of the plan. The first step towards developing a DM of a DISPLAN is codifying DM knowledge in DISPLAN, then later its reuse and sharing are facilitated. However, analysing the knowledge in a complex domain, such as DM, is not only difficult but also time-consuming. This paper continues the work that began in (Inan et al., 2015) to address the challenge of how to convert existing DM knowledge into constructs that can be easier to centrally store and share as required.

DM modelling generally aims to capture the complex characteristics of DM and present it in a way common people can understand easily (Sackmann et al., 2013). DM has these characteristics: a) Situatedness in an environment (Cavallo & Ireland, 2014). As disasters are dynamic, unpredictable and uncertain, an environment changes rapidly which can lead to the second characteristic. b) Time sensitivity (Janssen et al., 2010). In a disaster, every activity has to deal with deadlines, otherwise the consequences might lead to casualties (including fatalities) (Opper et al., 2010). c) Non-deterministic (Wex et al., 2014). Disasters often throw up unexpected eventualities. This factor means the unpredictability is very high. d) Presence of autonomous entities (Ernstsen & Villanger, 2014). This means that in a DM activity, individuals/agencies/organisations are coming from different backgrounds, knowledge, abilities, structure, mandate, no common perception and so on. The Agent-Oriented Analysis (AOA) approach enables analysis of complex systems, in particular socio-technical systems (Sterling & Taveter, 2009; Tran et al., 2007; Xu et al. 2011). AOA is easy to understand for humans, as it uses constructs from familiar organisational setting (e.g. roles, activities, interaction and etc.) (Miller et al., 2014). It is at a high level of abstraction that enables analysts to apply concepts from their daily deductive processes with which they are familiar (Winikoff & Padgham, 2013). There are clear similarities between the AOA and the context of DM, most strikingly: agents are driven by local goals and need to interact towards a system goal; agents have specified roles and need to interact accordingly; agents are situated and need to respond in real time in many instances (Lopez-Lorca et al., 2011). Not surprisingly, there have been various attempts recently to use AOA to support DM, for example, (Aldewereld et al., 2011; García-Magariño & Gutiérrez, 2013; Padgham et al., 2014; Scerri et al., 2012). However, much of these works focus on developing simulations of disaster events to gauge the effectiveness of existing practices.

A particular DISPLAN represents a set of goals stating the objectives of Disaster Management (DM) across each phase of the PPRR (Prevention, Preparedness, Response and Recovery) framework. Participants will be located in specific areas of authority and have hierarchical levels of control and command such as that of the NSW SES. Analysis and sharing of the knowledge involved requires a systematic approach to structure the knowledge and communicate it easily (Bera et al., 2011; Beydoun and Hoffmann 1998). The analysis requires answering complex questions such as: how a goal can be identified and evaluated; how agents negotiate their priorities as they collaborate in pursuit of common goal(s); what specific activities agents perform as they pursue their goal(s); what resources are needed for given goals or agents; what time and resource constraints should be imposed on particular agents; and so on. This paper applies AOA templates to convert disaster management to an intermediate form which can then be converted to DMM-based constructs. However, we note that AOA templates can be tuned to the analysis of DM knowledge by focussing on elements that are in common across various DISPLANs. In particular, commonalities are opportunely defined in *DISPLAN templates*. This work began in (Inan et al., 2015), where we developed an agent based process to facilitate the conversion of DISPLANS into a unified language, DMM (Othman et al., 2014). In this paper, we tune the process and validate it in a case study involving SES NSW DISPLANs. The DMM has been developed as a representative set of concepts and relations in a DM which can be used to store DM knowledge from real DM activities (Othman & Beydoun, 2013). DMM represents all concepts and relations in a DM domain. These concepts and relations were extracted from a collection of 98 DM models developed by various agencies including governmental, private sector and academia. Common constructs and rules from those models were reconciled and validated to produce DMM (Othman et al., 2014, p. 28).

## 3　Knowledge Analysis Framework

In this research, the analysis begins with the DISPLAN knowledge template, rather than the localised instance of the plan itself. The analysed DISPLAN document is in a semi-structured format which contains more data and information than unstructured one (Cooper et al., 2001). While the knowledge in that particular format is considered richer, it tends to be out lined in a business specification format (Selway et al., 2015) that hinders it to be adopted. Therefore, given that benefit of the document's structure, this research limits the input only to it before analysing and extracting it into to the structured one for both human and machine readable in the later development phase, for instance in an XML format. The use of templates as the input instead of the local DSIPLAN increases the effectiveness and efficiency of the analysis by first tuning the agent based model templates to suit the core structure of all DISPLANs. In this context, effectiveness relates to the adoption of the process in which the modellers producing the customised AB models are able to more quickly generate many instances of DISPLAN that are strongly based on the core template but are specific to localised parameters. This mirrors the approach taken by emergency management agencies. Further, templates are a benefit if any ratification of changes or updates occur as these can be promulgated and adapted in any instance of localised plans. Finally, templating is a key approach to effective interoperability as it helps stakeholders to quickly identify the urgent and relevant knowledge to respond to a particular activity when a participant is 'out of area' (from anther jurisdiction) by developing a familiar construct of actions which can easily be navigated. The application of the metamodel and customised AB models extends all of these efficiencies even further. The customised AB models generated from this process can be further adjusted according to hazard, context and resources where they will be implemented and will facilitate the ability to promulgate template changes across a digital repository of DISPLAN in real time. This is illustrated in Figure 1. In the case of a State level DISPLAN, the template can be employed to generate the plans for all municipalities/cities across the State, as they are all under the same hierarchy level. Therefore, all instances automatically conform to their template. For instance, in NSW, Australia, all the cities and regions across the State adopt the same DISPLAN template for flood disaster developed by the NSW SES (SES NSW, 2010). The template is developed as a classifier which is used by the NSW SES in each region and its cities to instantiate their specific DISPLANs. These particular DISPLANs adapt and adjust the customised template based on their resources and environments. In a similar vein, this can also be observed in the State of Victoria, Australia, for similar disasters (SES Victoria, 2011).

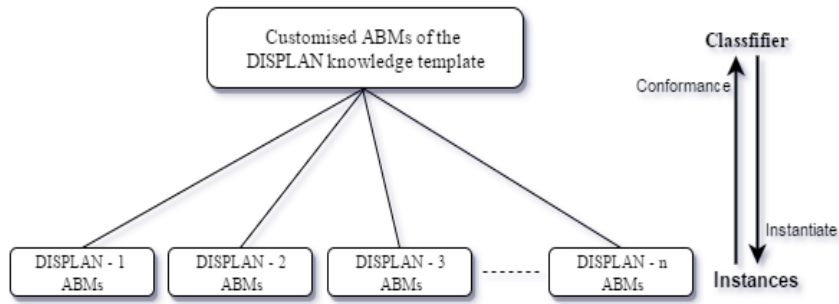

*Figure.1 The DISPLAN knowledge template as a classifier to instantiate*

As expected from the above discussion, in the first stage of our knowledge analysis framework, the knowledge engineer customises AB models with respect the DISPLAN template. The modeller is then able to synthesise and adjust them with respect to the environment and local resources of that city/municipality. The synthesised templates are then transformed into the repository following a specified semantic mapping. The knowledge structured in the repository can then be adopted by the particular city as its DISPLAN and shared and reused by other users for their DM activities. The knowledge analysis framework (shown in in Figure 2), thus consists of three stages, as follows:

Stage 1: the input is customised by seven Agent-Based models (ABMs) that are tightly coupled with the MOF. The input is the DISPLAN knowledge template across all PPRR phases in a semi-structured format. This process results in the customised AB models of DISPLAN knowledge templates.

Stage 2: the output of the Stage 1 will be handled as the input. The customised AB models of the DISPLAN knowledge template serves to generate the particular DISPLANs based on the specific local resources and circumstances. This process results in the AB model DISPLANs. In this stage, the repository is also prepared by annotating it. This produces an annotated DMM-based repository that is ready to be used for transformation processes.

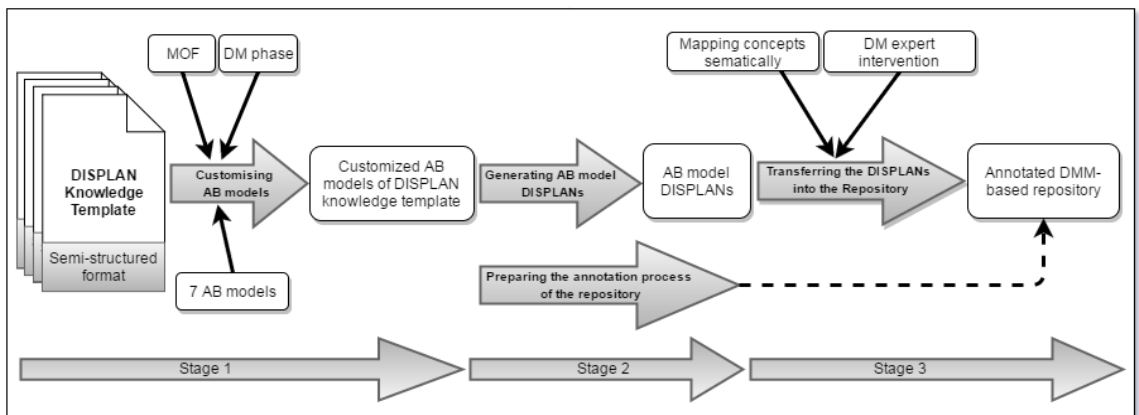

*Figure 2. Three main steps of the knowledge analysis framework*

Stage 3: This is the knowledge transformation process. It requires that the repository is in place and ready for the depositing processes. In this stage, the AB model DISPLANs produced in the second stage are transferred to the annotated DMM-based repository. A DM expert intervention is normally required to guarantee that the models resulting from the previous stage are mapped and positioned correctly to the appropriate concepts based on the semantic meaning.

The remainder of this section details each of the above stages of our knowledge analysis framework.

### 3.1  Stage 1: Customising Agent-Based Models

Agent Based Models (ABMs) can represent organisational processes and activities as described in a typical DISPLAN. The DISPLAN knowledge template in turn describes the structure of every DISPLAN. It also has knowledge that is common to all plans, for example contact details within the state or the names of roles. The template is in a semi-structured format and covers all four PPRR phases. In this step, the commonalities captured and expressed in the template is transferred to the agent based templates. That is, each agent based template undergoes three steps in this customisation:

1. Common knowledge elements are transferred to the ABMs, 2. each ABM template is reduced in size to delete elements that are not required, 3. Each element in the model is marked as either M0 or M1 (this acts as a pointer in the transfer in stage 3). That is, only the required elements are used in the ABMs. Essentially, this process is to analyse and extract any feature characteristics of the DM knowledge in the document and structure them in their corresponding ABMs. The output of this stage is the customised AB models of DISPLAN knowledge templates.

We identify the following set of seven AO templates to capture the knowledge from DM. The details of these models are based on (Lopez-Lorca et al., 2016) and are as follows:

1. A *goal model* represents a particular goal that an agent persistently strives to accomplish. It describes the proactiveness of an agent. The model consists of the main goal, sub-goal(s) and role(s). The main goal is the goal that needs to be achieved by a set of activities represented as the sub-goals. At least one role will be responsible for each goal (main and/or sub-goal) in the model.
2. A *role model* represents all knowledge responsibilities that need to be played by each of the involved agents. The knowledge basically represents the same knowledge as that in the *goal model*. The constraints of each agent playing any particular role are also listed, if available.
3. An *organisation model* is used to represent the relationships between agents playing roles, and highlights how to take in to account their relationships in a DM process. The model defines the communication channels between agents in different organisations or levels of command in a widely dispersed disaster. These agent responsibilities are the same as in the *goal model*.
4. An *interaction model* elaborates on the extent to which agents playing roles interact each other. In the DM activity, this model lays the groundwork for two or more agents interacting with each other, based on specific purposes.
5. An *environment model* specifies the environmental knowledge and resources that agents playing roles will require in the DM activities.
6. An *agent model* describes how particular agents are involved in DM activities, along with further specifications of their activities and their goals. The triggers identified in this model represent the reactiveness of the external event(s) that spur agents into action, as agents are situated in that environment.
7. A *scenario model* binds all knowledge elements as activities that need to be undertaken in pursuing a particular goal. The activities are preceded by a pre-condition and followed a post-condition, as a desired state of the goal being pursued in the activities. These activities are specified as either parallel, sequential, or interleaved.

### 3.2 Stage 2: Preparing the knowledge transfer process

As can be seen in Figure 2, essentially there are two activities in this stage. They are: generating the AB model's DISPLAN, and preparing the repository by annotating it. As discussed previously, in the first stage, the process is to produce the customisable AB model of the DISPLAN knowledge template. This will serve as a classifier DISPLAN template that can instantiate the particular DISPLANs out of it, before transferring them in to the prepared repository. The two subsections are explained as follows:

#### 3.2.1 Generating Agent-Based Model DISPLANs

The DM knowledge deposited in the repository is the one that can be utilised directly by the stakeholders. Knowledge is specified and related to a certain characteristic which people on the ground can understand directly and contextually. Therefore, it is important to ensure that before transferring into the repository, the knowledge has been generated to have a particular characteristic. For instance, if the process is aimed to produce a DISPLAN for a municipality "ABC" under the region "PQR", then all the general characteristics of the classifier need to be narrowed down to the specific one representing the municipality "ABC". The AB model DISPLAN "ABC" resulted in this stage automatically conforms to all the general variables from its template which subsequently will be adjusted with the local ones. The ABM DISPLAN is now ready to be transferred.

#### 3.2.2 Annotating DMM concepts with the AO concepts

This step prepares the structure of the repository to receive the ABMs from Stage 1. The repository structure is based on the DMM. This step provides the basis of a mapping between the elements of the agent based models to DMM constructs. We use a corresponding Agent-Based, FAML (Beydoun et al., 2009; Beydoun et al. 2006), which describes agent models at MOF abstractions, to annotate DMM with agent based tags. This annotation process is a one-off process for all concepts in the DMM with their appropriate agent tags. Table 1 shows examples DMM concepts and their corresponding tags.

| DMM Concept | Agent based tags | Description |
|---|---|---|
| Preparedness Goal | <<Goal>> | Represents a certain condition needed to be achieved by the system |
| PreparednessTask | <<Role>> | Represents a set of capabilities to be performed by agent to achieve the goal(s) |
| PreparednessTeam | <<Agent>> | Represents an entity that, having certain properties, can play one or more role(s) |
| Training & Public Education | <<Activity>> | Describes a set of activities to be performed to achieve the goal(s) |
| Before-disaster | <<Event>> | Defines a situation change that influences a significant change of an agent to respond to the situation |
| Media & MutualAidAgreement | <<EnvironmentEntity>> | Represents any resources required to perform the tasks |

*Table 1. Examples of AOSE metamodel Concepts mapped to DMM concepts*

The DMM has 92 concepts: 21 concepts in Prevention, 25 concepts in Preparedness, 25 concepts in Response and the other 21 concepts in Recovery. More than one concept in the DMM can get mapped to the same agent based concepts. For example, many concepts in the DMM are about activities and resources/supporting systems. Hence, agent concepts of <<Activity>> and <<EnvironmentEntity>> are repeatedly mapped to many concepts in the DMM. From Table 1, *Training* and *PublicEducation* concepts are annotated as <<*Activity*>>, *MutualAidAgreement* and *Media* are annotated as <<*EnvironmentEntity*>>. Whilst this process is undertaken only once, the knowledge modeller can always revisit to improve the mapping process if required. Figures 3 shows the annotated DMM concepts in Preparedness.

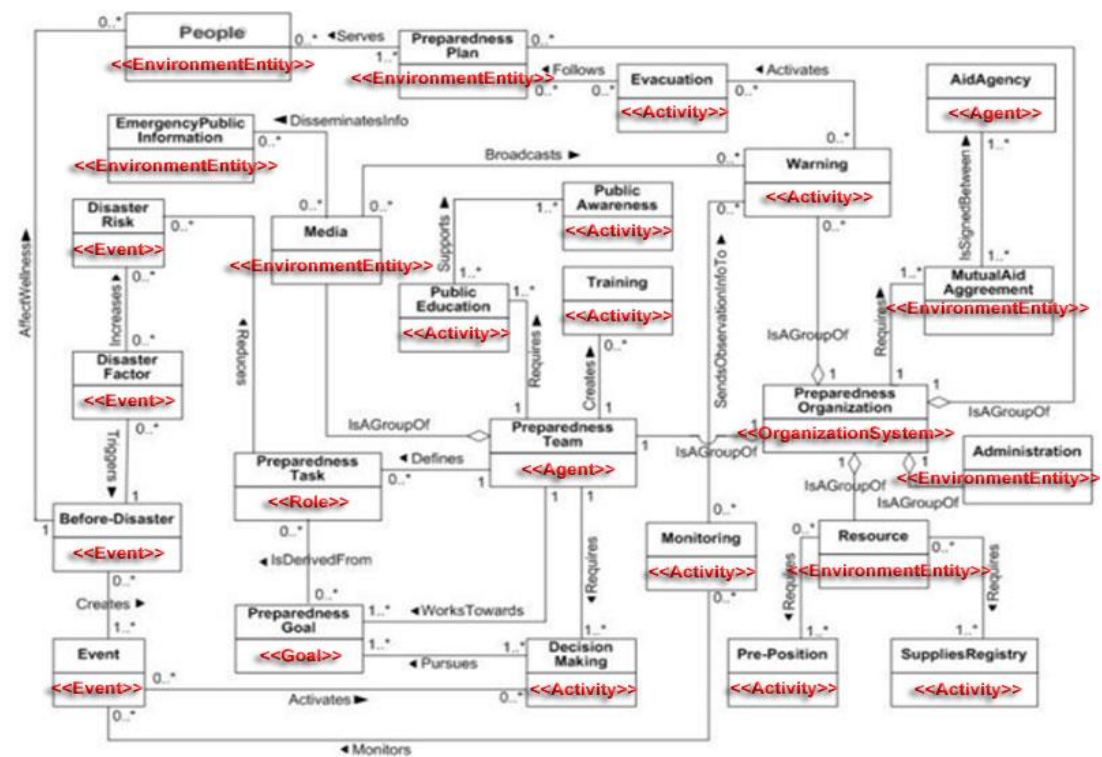

*Figure 3 Preparedness-phase of DMM annotated with AO concepts*

## 3.3  Stage 3: Transferring ABM of DISPLANs into a DMM-based repository

In this stage, the ABM DISPLANs acquired from the Stage 1 are transferred into the annotated DMM representation using the mapping process mechanism described above. This process is semi-automatic. A DM practitioner is involved in transferring each of the AB models to each of their appropriate annotated DMM constructs. There are seven corresponding AB models that need to be mapped to 92 DMM concepts across all phases in the DM. The DM practitioner is involved by judging semantically each of the concepts in the DMM that is appropriate to capture the information in the each of the AB knowledge models. The DM practitioner involvement in this mapping mechanism is recognition of the fact that that human capability for decision-making system cannot be fully replaced. With respect to the MOF hierarchy, not all ABMs are represented equally. Some models generate more constructs at M0 level than others. Other models generate more constructs a M1 level. For instance, Scenario and Agent models generate more constructs at M0, while Role and Goal models generate more constructs at M1 level. The overall process is validated in converting NSW State Emergency Services DISPLAN templates to DMM constructs. A DISPLAN instance that the research focuses on is the Wagga Wagga, Murrumbidgee Valley in NSW. A DM practitioner previously from the SES NSW (the third author) is involved in this process. This case study is described in the next section.

## 4 Case study: Customising Agent-Based Flood DISPLAN knowledge of Wagga Wagga City, NSW

The Wagga Wagga Local Flood Plan (LFP) is a flood hazard specific DISPLAN prepared by NSW SES in collaboration with the Wagga City Council and a Local Emergency Committee. The LFP is a sub plan/supporting plan to the NSW SES State Flood Plan. The LFP also supports a Regional Emergency Management Plan (EMPLAN) prepared by NSW Police and regional government agencies. Other plans focusing on Health, Agriculture and Energy and Utilities etc., also support the Regional EMPLAN and LFP and are enacted during disasters such as floods. The Regional EMPLAN is in turn a sub plan supporting the NSW State Emergency Management Plan (State EMPLAN). The Wagga Wagga LFP is maintained to prepare for, manage the response to and support recovery from flood disasters. The LFP is a semi-structured document, written in a particular style and structured by practitioners involved in the DM for floods. It covers knowledge in three phases: Preparedness, Response and Recovery. In this paper, the knowledge analysis framework is illustrated in the Preparedness and Response phases for the roles and responsibilities of the stakeholders. The details of this are presented in this section.

### 4.1 Stage 1: Customising AB models

A *goal model* is generated as a case point (Figure 4). The elements shown between "<" and ">" are common to all DISPLANs. In addition, the various elements are tagged by either M0 or M1. These refer to MOF abstraction levels, and will serve to disentangle the fuzziness of concepts as they are transferred later in the repository.

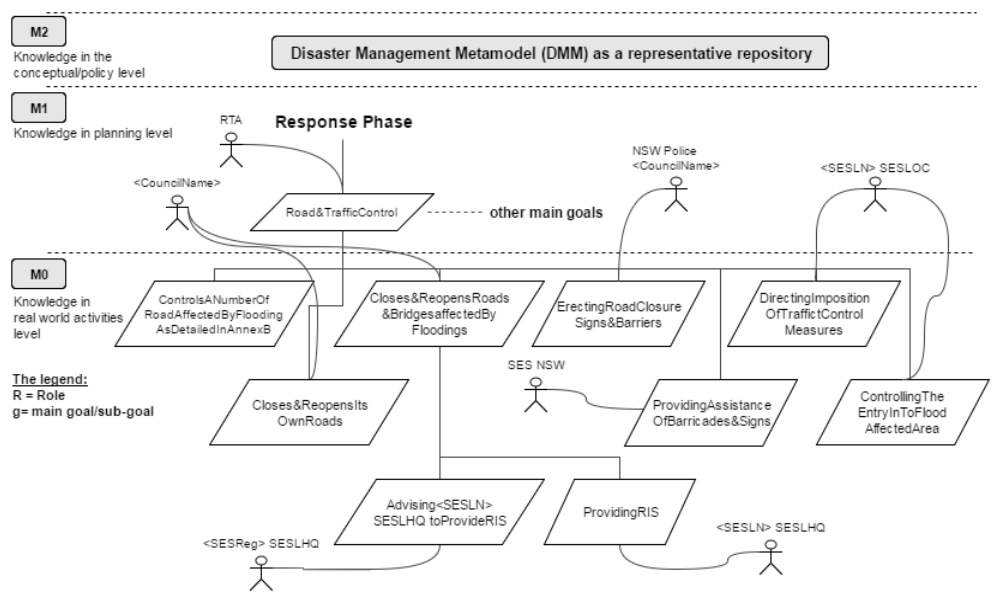

*Figure 4 Customizing the goal model, an example*

In the *goal model*, it can be identified that there is at least one role responsible for each of the goal (main goal and its sub-goals). For instance, for the sub-goal: *Providing Road Information Service (RIS)*, the role *<SESLN> SESLHQ* (Local Headquarters) is responsible. In addition, this *goal model* structure also shows that not only is the *<SESLN> SESLHQ* responsible for that particular goal, but *<CouncilName>* and *Road and Traffic Authority (RTA)* (now RMS) will be interacting in pursuing the sub-goal. These customizing processes apply to the rest of the DISPLAN knowledge template. Eventually a complete customized *goal model* of the DISPLAN knowledge template is structured. As for the other ABMs, the same process is applied iteratively. The other six ABMs (*role model, organisation model, interaction model, environment model, agent model* and *scenario model*) of DISPLAN knowledge template are produced following this approach.

## 4.2  Stage 2: Generating Agent-Based Model DISPLANs

In this stage, the customised ABM of flood DISPLAN knowledge templates produced in Stage 1 generate the Wagga Wagga ABM DISPLAN knowledge. All the knowledge element classes in the templates instantiate the specific elements representing the Wagga-Wagga municipality. The instance conforms and inherit all the knowledge elements of the template. This is ensured by substituting the class of knowledge elements of the template with the corresponding ones in the ABM model template. e.g. Figure 5 shows the knowledge classes conforming with the template shown in Figure 4 earlier.

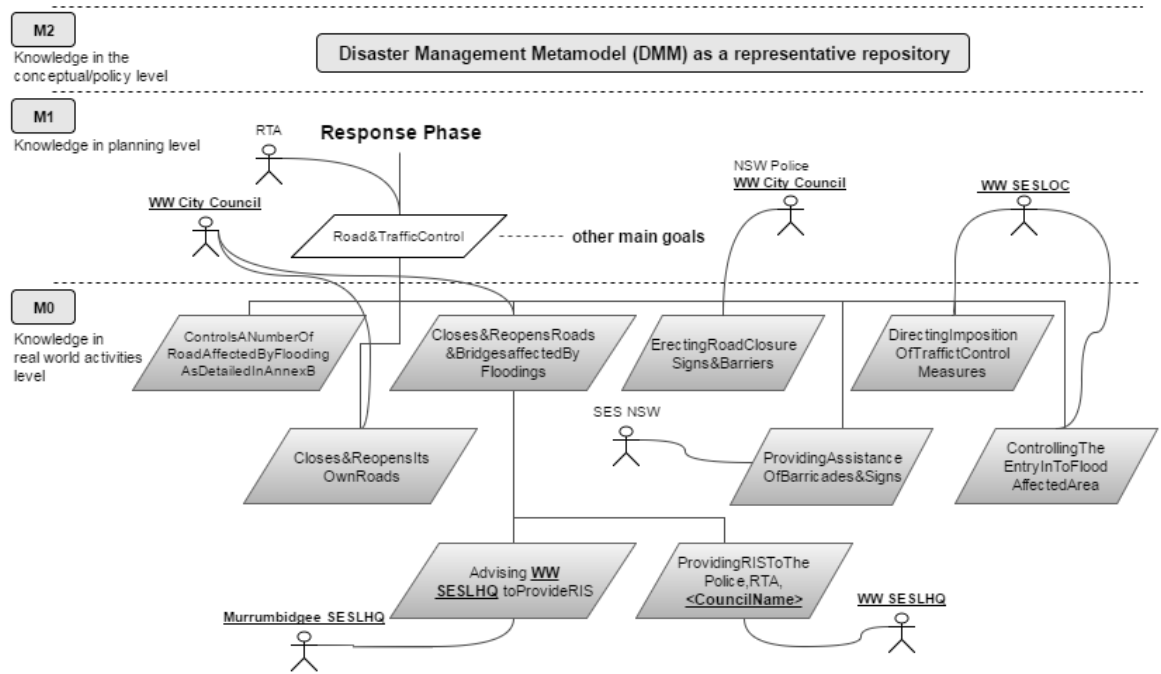

Figure 5 *The goal model of Wagga-Wagga DISPLAN knowledge generated from the template. All the roles generated from the template are underlined and the sub-goals are highlighted.*

The process also applies the other models. Finally, the ABMs of Wagga-Wagga (WW) DISPLAN knowledge is ready to be transferred to the repository.

## 4.3  Stage 3: Transferring ABM of DISPLANs into DMM-based repository

A database system is employed, designed to assist the stakeholders to identify their specific roles and goals for a specific disaster phase and action e.g. road closure in the response phase, by displaying a customised view of only the relevant knowledge and interactions with other stakeholders and concepts. The system will automatically limit and focus appropriate annotated DMM concepts from the DMM. In other words an emergency manager uses a concept driven interface to select a disaster management task in a specific phase of the disaster and is automatically shown: 1. their goal and sub goals, 2: their role and responsibilities 3: who they should interact with in an organisational (inter and intra agency) context and the communication channels to interact through, 4: the purpose of each interaction with another stakeholder, 5: the environment or parameters in which they are likely to operate in and which will impact their actions, 6: the triggers of when they are required to act and perform their role as an agent in the process, and 7: an overarching scenario binding all the elements together and indicating what the pre-condition and post conditions will be. This assists the DM practitioner in having broad awareness but concise visibility of the related knowledge, processes and concepts that they must apply within the DISPLAN and the overarching DMM. This process is applied in all mapping activities between all ABMs and the corresponding annotated concepts of the DMM metamodel. Thus, the DM practitioner, no matter what role they perform at any level within the incident management team is able to refocus the information on any action or context within the plan and will have a much clearer understanding of their responsibilities in the DISPLAN. In the repository, the three components: DM phases, the MOF framework and the ABMs construct the knowledge in a three-dimensional (3D) structure which allows the knowledge to be drilled down or rolled up easily at in real time during the DM activities. This is illustrated in Figure 6 where the *goal model* of Wagga-Wagga DISPLAN knowledge of the Response phase is transferred to the concept in the repository that responsible for the *goal model,* that is the <<goal>>. Therefore, only the <<goal>> of the Response phase is highlighted for all MOF layers. A DM expert intervenes the process by mapping the process semantically. The stakeholders can then holistically pin point the appropriate knowledge through each cube of the structure. To complete the 3D knowledge structure, these three stages are undertaken iteratively. Eventually, this allows the knowledge to be shared and reused easily.

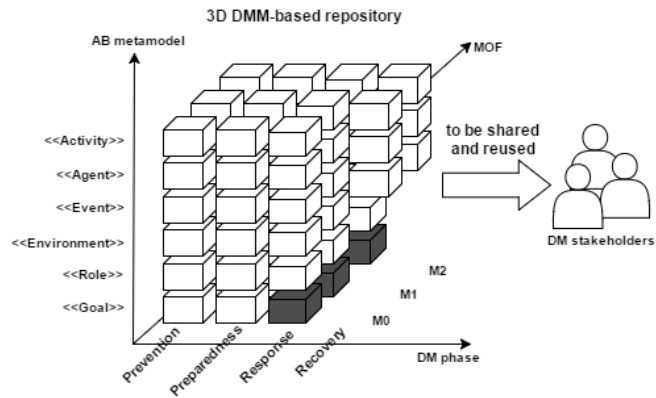

*Figure 6 Three-dimensional knowledge structure in the annotated DMM-based repository*

This process is implemented as a graphical web-based user interface that supports access to the DMM knowledge repository. The ABMs are made available in XML and they are input to the database. The database type used is MySQL as it is a powerful, widely used as well as an open source database; it also harnesses the Apache web-based technology to connect client requests to the server. This web server technology is used in the most web servers around the world.

## 5  Summary and Future Research

The conversion to the unifying DM metamodel is underpinned by an Agent Oriented Analysis process. The process is tuned to the role and responsibilities elements of DISPLANs, by first identifying commonalities across plans and adjusting the Agent Based Model (ABM) templates. The DISPLANs templates are used in this tuning phase. The DISPLANs templates are typically used by well-established emergency management authorities (e.g. NSW SES or VIC SES). The DM metamodel constructs are applied to the subsequently generated agent-based models that represent the DISPLANs. This is supported by a semi-automated step to guide a DM practitioner on how to use the metamodel constructs. A tool is developed to support the process is described in the paper. A DM practitioner validated the conversion process from DISPLANs to ABMs to the metamodel-based repository. With the use of the tuning step, the conversion process was significantly quicker. The approach was illustrated using the actual flood DISPLAN knowledge template used by SES NSW to generate a DISPLAN of the Wagga Wagga City Council area on the Murrumbidgee River in NSW. The paper illustrated the effectiveness of the overall mapping the DISPLAN to the unifying metamodel, DMM.

The Design Science Research (DSR) approach (Hevner et al., 2004) of Information System adopted in this research contributes to the development of knowledge analysis framework as the built artefacts (Gregor & Hevner, 2013). To conclude, this research contributes in (1) Justifying that ABMs from Software Engineering is considered as the most representative methodology to capture the complex knowledge characteristics of the DM activities; (2) Defining that the analysed and structured DM knowledge are disentangled by adopting the MOF specified by OMG; and the last (3) Defining the knowledge transfer process to the DMM-based repository semantically which has been annotated previously. As in the DSR, once built, the evaluation of the artefacts is conducted at the first place (Hevner, 2007) by following this (Pries-Heje et al., 2008). Adopting this evaluation strategy (Prat et al., 2014), the research has successfully demonstrated that the use of template has not only the significant impact of the framework effectivity and efficiency but also this can boost the DM resilience endeavours by helping DM authoritative agencies to developed local wisdom-based DISPLANs.

The work successfully generates an actual DISPLAN from the template and maps it into the metamodel representation. Future work will further develop the process by engaging further DM collaborators to externally validate the efficacy of the developed framework against various validity threats. This will then be followed by an evaluation of the effectiveness of the repository from the practitioner's perspectives. While there are multiple significant issues for emergency managers which can benefit from the work outlined in this paper, for example improving the inefficient maintenance of such a large connected but disparate knowledge representation currently maintained as individual documents, or increasing the effectiveness and real time use of response plans by curating focused role and action based incident decision process, the critical outcome discussed in this paper is the importance of shared understanding in the planning process and ease of access to disaster management knowledge, roles and actions.